\documentclass[11pt]{article}
\usepackage{amsthm}
\usepackage[title,page]{appendix}
\usepackage{myMac-sig-alt}
\draftdim
	\newboolean{FIG_FLAG}
\setboolean{FIG_FLAG}{true} 
\setboolean{FIG_FLAG}{false}
	\newcommand{\figchoose}[2]{\ifthenelse{\boolean{FIG_FLAG}}{
	  #1}{#2}}
	\setboolean{FIG_FLAG}{true}	
	\setboolean{FIG_FLAG}{false}	

	\newenvironment{mini_itemize}{\begin{list}%
	  {$\bullet$}%
	  {\setlength{\parsep}{0ex}%
	   \setlength{\itemsep}{0.5\parskip}%
	   \setlength{\topsep}{-0.5\parskip}%
	}}{\end{list}}

	\newcommand{\ulx}{\underline{x}}

	\newcommand{\htw}{\hat{w}}
	
	\newcommand{\cen}{\mathrm{cen}}
	
	\newcommand{\magn}{\mathrm{mag}}

	\newcommand{\via}{\mathrm{via}}
	\newcommand{\tmp}{\mathrm{tmp}}
	\renewcommand{\TT}{{\cal T}}
	\renewcommand{\SS}{{\cal S}}

	\newcommand{\JC}{\mathrm{JC}}
	\newcommand{\MK}{\mathrm{MK}}
	\newcommand{\Root}{\mathrm{Root}}
	\newcommand{\std}{\mathrm{Std}}

	\renewcommand{\myPara}[1]{
		\addtocounter{myParagraph}{1}
	       	\vspace*{-1.0\abovedisplayskip}
		\paragraph{\cocyan{\P \themyParagraph. #1}}
		}

\begin{document}
\title{Isotopic Arrangement of Simple Curves:\\
	an Exact Numerical Approach based on Subdivision}
\author{
	\Large{Jyh-Ming Lien}\\
	Department of Computer Science\\
	George Mason University \\ Fairfax, VA 22030, USA\\ jmlien@cs.gmu.edu\\
        \vspace*{0.5cm}\\
        \Large{Vikram Sharma}\\
        The Institute of Mathematical Sciences, HBNI, Chennai, India.\\
        {vikram@imsc.res.in}\\
        \vspace*{0.5cm}\\
      \Large{Gert Vegter}\\
        Johann Bernoulli Institute for Mathematics and Computer Science\\
        Nijenborgh 9,   9747 AG Groningen\\
        The Netherlands.\\
        gert@rug.nl\\
        \vspace*{0.5cm}\\
       \Large{Chee Yap}\\
        Department of Computer Science\\ Courant Institute of Mathematical Sciences\\
        New York University. New York, USA.\\
        yap@cs.nyu.edu}
\date{}
\maketitle
\thispagestyle{empty}
\begin{abstract}
This paper presents the first purely numerical (i.e., non-algebraic)
subdivision algorithm for the isotopic approximation
of a simple arrangement of curves.
The arrangement is ``simple'' in the sense that 
any three curves have no common intersection,
any two curves intersect transversally,
and each curve is non-singular.
A curve is given as the zero set of an analytic function $f:\RR^2\to\RR^2$,
and effective interval forms of $f, \pdiff{f}{x}, \pdiff{f}{y}$
are available.  Our solution generalizes the
isotopic curve approximation algorithms of Plantinga-Vegter (2004)
and Lin-Yap (2009).  

We use certified numerical primitives based on interval methods.
Such algorithms have many favorable properties: they are practical,
easy to implement, suffer no implementation gaps,
integrate topological with geometric computation,
and have adaptive as well as local complexity.  

A version of this paper without the appendices appeared in \cite{lien-sharma-vegter-yap:arrange:14}.
\end{abstract}

\newpage
\setcounter{page}{1}
\section{Introduction}

We address problems in computing approximations to curves and surfaces.
Most algebraic algorithms for curve approximation begin by computing
a combinatorial object $K$ first.   To compute $K$, we typically use 
algebraic projection (i.e., resultant computation), followed by root isolation
and lifting.  But most applications will also require the geometric realization $G$.
Thus we will need a separate (numerical) algorithm to compute $G$. 
This aspect is typically not considered by algebraic algorithms.

In this paper, we describe a new approach for computing curve arrangements
based on purely numerical (i.e., non-algebraic) primitives.  
Our approach will integrate the computation of the
combinatorial ($K$) and geometric ($G$) parts.
This leads to simpler implementation.
Our numerical primitives are designed to work directly
with arbitrary precision dyadic (BigFloat) numbers,
avoiding any ``implementation gap'' that may mar abstract algorithms.
Furthermore, machine arithmetic can be used as long as no over-/underflow occurs,
and thus they can serve as efficient filters \cite{bbp:interval-filter:01}.

We now explain our specific problem, and illustrate the preceding notions of $K$ and $G$.
By a \dt{simple curve arrangement} we mean a collection
of non-singular curves such that no three of them intersect,
and any two of them intersect transversally.
The simple arrangement of three or more curves can, in some sense,
be reduced to the case of two curves (see the Final Remarks).  Let
$ F : \RR^2\to\RR^2$, where $F(x,y) = (f(x,y), g(x,y))$
is a pair of analytic functions.   It generically defines two planar curves
$S=f^{-1}(0)\ib\RR^2$ and $T=g^{-1}(0)$. We call $F=0$ a \dt{simple system} of equations if 
$\set{S,T}$ is a simple curve arrangement.
Throughout this paper, $F=(f,g)$ will be fixed unless otherwise indicated.
\refFig{arrangement} illustrates such an arrangement for the curves defined by
$f(x,y)=y-x^2$ and $g(x,y)=x^2+y^2-1$.
The concept of hyperplane arrangement is highly classical in computational geometry \cite{bkos:bk}.
Recent interest focuses on nonlinear arrangements \cite{boissonnat-teillaud:bk-06}.

\figchoose{
\FigEPS{arrangement}{0.7}{Arrangement of two curves, $y=x^2$ and $x^2+y^2=1$}}
{\vfigpdf{Arrangement of two curves, $y=x^2$ and $x^2+y^2=1$}{arrangement}{0.7}}

Our basic problem is the following:
suppose we are given an $\eps>0$ and a region $B_0\ib \RR^2$,   
called the \dt{region-of-interest} or ROI, which is usually in the shape of an axes-aligned box.
We want to compute {\em an $\eps$-approximation to the arrangement
of the pair $(S, T)$ of curves restricted to $B_0$.}
This will be a planar straightline graph $G=(V,E)$
where $V$ is a finite set of points in $B_0$
and $E$ is a set of polygonal paths in $B_0$.
Each path $e\in V$ connects a pair of points in $V$, and
no path intersects another path or any point in $V$ (except at endpoints).
Moreover, $E$ is partitioned into two sets $E=E_S\cup E_T$
such that $\cup E_T$ (resp., $\cup E_S$) is an approximation of $T$ (resp., $S$).
The correctness of this graph $G$ has two aspects:
(A) topological correctness, and (B) geometric correctness.
Geometric correctness (B) is easy to formulate:
it requires that the set $\cup E_S \ib B_0$ is $\eps$-close to $S$
in the sense of Hausdorff distance: $d_H(S,\cup E_S)\le \eps$.
Similarly, the $\cup E_T$ is $\eps$-close to $T$.
If we specify $\eps=\infty$, then we are basically unconcerned
about geometric closeness.

Topological correctness (A) is harder to capture.
One definition is based on the notion of ``cell decomposition''.
A (cell) \dt{decomposition} of $B_0$ is a partition $K^*$ of $B_0$ into a collection
of sets called cells, each $c^*\in K^*$ homeomorphic to a closed $i$-dimensional ball
($i\in\set{0,1,2}$); we call $c^*$ an $i$-cell and its dimension is $\dim(c^*)=i$.
If $b^*$ is an $i$-cell and $c^*$ an $(i+1)$-cell, we say
$b^*$ \dt{bounds} $c^*$ if $b^*$ is contained in the boundary $\partial c^*$ of $c^*$.   
Call $K^*$ an $(S,T)$-decomposition of $B_0$ if the set $(S\cup T)\cap B_0$
is a union of some subset of $0$- and $1$-cells of $K^*$.
A $(S,T)$-decomposition is illustrated in \refFig{arrangement}(b).

A \dt{cell complex} $K$ 
is an (abstract) set such that each $c\in K$ has a specified $\dim(c)\in \set{0,1,2}$
together with a binary relation $B\ib K\times K$ such that $(b,c)\in B$ implies
$\dim(b)+1=\dim(c)$.   
We say that the decomposition $K^*$ is a \dt{realization} of $K$,
or $K$ is an \dt{abstraction} of $K^*$,
if there is a 1-1 correspondence between the cells $c^*$ of $K^*$ with the elements $c\in K$
such that $\dim(c^*)=\dim(c)$, and moreover the relation $(b,c)\in B$ iff
$b^*$ bounds $c^*$ in $K^*$.
\refFig{arrangement}(c) shows the abstraction $K$ of the
decomposition in \refFig{arrangement}(b).

Our algorithmic goal is to compute a planar straightline graph
(PSLG for short \cite{preparata-shamos}) $G=(V,E)$ which approximates $(S,T)$ in a box $B_0$.
Such a graph $G$ naturally determines a decomposition $K^*(G)$ of $B_0$ as follows:
the set of $0$-cells is $V$, the set of $1$-cells is $E$ and
the set of $2$-cells is simply the connected components of $B_0\setminus (V\cup (\bigcup E))$.
Finally, we say $G$ is topologically correct if there exists an $(S,T)$-decomposition
$K^*$ such that $K^*$ and $K^*(G)$ are realizations of the same abstract cell complex.

\myPara{Towards Numerical Computational Geometry.}
The overall agenda in this line of research
is to explore new modalities for designing geometric algorithms.
We are interested in exploiting weaker numerical
primitives that are only complete in a certain limiting sense.
Unlike traditional exact algorithms, our algorithms must
strongly interact with these weaker primitives, and exploit adaptivity.
The key challenge is to achieve the kind of exactness and guarantees
that is typically missing in numerical algorithms.  
See \cite{yap:praise:09} for a discussion of ``numerical computational geometry''.

\ignore{
	The modus operandi for designing algorithms in Computational Geometry (CG)
	is to first postulate some abstract geometric
	primitives, and then design the algorithm using these primitives.  These
	primitives include operators (e.g., intersecting two lines)
	and predicates (e.g., test if a point lies to the left of a line).  
	The advantage of such an approach is that the algorithm design
	and the implementation of the primitives are cleanly decoupled.
	The disadvantage is that such primitives can become a major
	issue in implementation.  A common response to implementing difficult
	primitives is to simply implement them in
	machine precision arithmetic, thereby losing any ``exactness'' associated with the original
	algorithm.   Of course, this leads to the nonrobustness problems
	that Computational Geometers attacked in the last two decades.
	The Exact Geometric Computation (EGC) solution (as represented by
	\leda, \cgal, or \corelib) is to ensure that the predicates are error-free.
	For example, Emiris and Karavales showed how the primitives
	in the construction of Apollonius diagrams (i.e., Voronoi diagram of circles)
	\cite{emiris-karavales:apollonius:06} can be efficiently implemented.
	Nevertheless, such algorithms are not easy to implement and hard to generalize.
	In many situations, the price of using exact primitives does not seem justified.
	E.g., a popular approach to approximating
	an implicit surface is to incrementally ``shoot rays'' to collect
	(algebraic) sample points on the surface.
	If these points are approximated, they are unlikely to lie of the surface.  Hence in
	the decoupled approach, one must find these algebraic sample points exactly, but this
	cost does not seem justified for a computation
	whose ultimate object is only an approximation. 
}
In the algebraic approach, one must compute the abstract complex $K$ before the
approximate embedded graph $G$.
Indeed, most algebraic algorithms do not fully address the computation of $G$.
In contrast to such a ``decoupled'' approach, our algorithm provides
an integrated approach whereby we can commence to compute $G$ (incrementally)
even before we know $K$ in its entirety.  Ultimately, we would be able
to determine $K$ exactly ---
this can be done using zero bounds as in \cite{yap:subdiv1:06,burr+3:subdiv2:12}.
The advantage here is that our integrated approach can cut off
this computation at any desired resolution, without fully resolving
all aspects of the topology.   This is useful in applications like visualization.

Unlike exact algebraic primitives, our use of analytic (numerical)
primitives means that our approach
is applicable to the much larger class of analytic curves.
Numerical algorithms are relatively
easy to implement and  have adaptive as well as ``local'' complexity.
Adaptive means that the worst case complexity does not characterize
the complexity for most inputs, and local means the computational effort
is restricted to ROI.

One disadvantage of our current method is that it
places some strong restrictions on the class of curve arrangements:
the curves must be non-singular with pairwise transversal intersections in the ROI.
In practice, these restrictions can be ameliorated in different ways.
The complete removal of such restrictions is a topic of great research interest.

The algorithms in this paper fall under the popular literature on Marching-cube type algorithms
\cite{mc-survey:06}.  There are many heuristic algorithms here which are widely used.
The input for these algorithms
can vary considerably.  E.g., Varadhan et al.\
\cite{varadhan+3:adaptive:04,varadhan+4:voxel:03}
discuss input functions $F:\RR^3\to\RR$ that might be a discretized function, or a CSG model or 
some polygonal model -- each assumption has its own exactness challenge.

\section{Our Approach: Isotopic Curves Arrangement}

All current exact algorithms for curve arrangements are based on algebraic projection,
i.e., they need some resultant computation.
The disadvantage of projection is the large number of
cells: even in relatively simple examples, the graph can be large as seen
as \refFig{arrangement}(c).  For many applications, the 2-cells 
may be omitted, but the graph remains large.  There are several known techniques
to reduce this (double-exponential in dimension) explosion in the number of cells.
In this paper, we avoid cell decomposition, but base our topological correctness
on the concept of isotopy.
Our algorithm uses the well-known subdivision paradigm, and produces
a subdivision of the input domain into boxes.
 \refFig{eg50} illustrates the form
of output from our subdivision algorithm using
our previous example of $y=x^2$ and $x^2+y^2=1$.\footnote{
The figure is not produced by the algorithm
of this paper because the implementation is currently underway.
Instead, it is produced by the Cxy Algorithm for approximating a non-singular curve
\cite{lin-yap:cxy:11}, using the input curve $fg=0$.   Thus the intersection points
are singularities which the Cxy algorithm cannot resolve, but this
does not prevent its computation to some cut-off bound.  Also, the
Cxy algorithm does not know which part of the arrangement is the $f$-curve
and which is the $g$-curve.
}
The number of subdivision boxes tend to be even more numerous than
cells in the decomposition approach. 
But these numbers are not directly comparable to number of cells for three reasons:
(1) Subdivision boxes are very cheap to generate.
(2) Most of these boxes can be instantly discarded as inessential
for the final output (we keep them for visualization purposes).
(3) Unlike cells, our subdivision boxes play a double role:
they are used for (A) topological determination as well as (B) in determining
geometric accuracy.

The approach of this paper has previously been successfully applied
to the isotopic approximation of a single non-singular curve or surface
by Plantinga and Vegter
\cite{plantinga-vegter:isotopic:04,plantinga:thesis:06}
and Lin and Yap \cite{lin-yap:cxy:11,lin:thesis}.
The current paper is a non-trivial extension of these previous works.


We now define the notion of isotopy for arrangements.
For our problem on arrangements, we need to extend the standard
definitions of isotopy.
Suppose $S,T\ib\RR^2$ are two closed sets and $\eps>0$.
First recall that $S$ and $T$ are (ambient) \dt{isotopic} if
there exists a continuous mapping
	\beql{gamma}
	\gamma: [0,1]\times\RR^2\to\RR^2
	\eeql
such that for each $t\in[0,1]$, the function $\gamma_t:\RR^2\to\RR^2$
(with $\gamma_t(x,y)=\gamma(t,x,y)$) is a homeomorphism,  $\gamma_0$ is the identity map,
and $\gamma_1(S)=T$.  
If, in addition, $d_H(S,T)\le \eps$ (where $d_H$
is the Hausdorff distance on closed sets)  we say that they are \dt{$\eps$-isotopic}.
We will write
	$$S\stackrel{\eps}{\simeq} T ~(\via~\gamma)$$
in this case.  Note that we may omit mention of $\eps$, in which case
it is assumed that $\eps=\infty$. 

We now generalize this to arrangement of sets.
Let $\ol{S}=(S_1\dd S_m)$ and $\ol{T}=(T_1\dd T_m)$ be two sequences of $m$ closed sets.
For each non-empty subset $J\ib \set{1,2\dd m}$, let $\ol{S}_J$ denote the
intersection $\cap_{i\in J} S_i$.  Similarly for $\ol{T}_J$.
We say that $\ol{S}$ and $\ol{T}$ are \dt{isotopic} if 
there exists a continuous mapping $\gamma$ as in \refeq{gamma} such that
for each non-empty subset $J\ib \set{1,2\dd m}$, we have
	$$\ol{S}_J \stackrel{\eps}{\simeq} \ol{T}_J ~(\via~\gamma).$$
We also call $\gamma$ an \dt{isotopy} from $\ol{S}$ to $\ol{T}$.
For simple curve arrangements, the critical problem to solve
is the case $m=2$.   We assume the
two curves $S_1,S_2$ are restricted to a region or box $B$.
Our basic problem is to compute a pair of curves $(T_1,T_2)$ such that
	\beql{iso}
	(T_1,T_2)\stackrel{\eps}{\simeq} (S_1\cap B, S_2\cap B).
	\eeql
The approximations $(T_1,T_2)$ produced by our algorithms will be piecewise linear curves.
See \cite{boissonnat+4:mesh:06} for a general discussion of isotopy of the case $m=1$.

\subsection{Normalization relative to a Subdivision Tree}

In Appendix A, we provide the necessary definitions; these are
consistent with the terminology in the related work \cite{lin-yap:cxy:11}.
For now, we rely on common terms that are mostly self-explanatory.

\myPara{Box Complexes and Subdivision Trees.}
Our fundamental data structure is
a \dt{subdivision tree} $\TT$ rooted in some box $B_0$.
In 2-D, $\TT$ is the well-known quad-tree and $B_0$ is a rectangle.
Each internal node of $\TT$ has four congruent children.
The boxes of a subdivision tree are non-degenerate (i.e., $2$-dimensional).
They need not be squares, but for the correctness of our algorithm, their aspect ratios
must be $\le 2$.
For any region $R\ib\RR^2$, we define a \dt{subdivision} of $R$ to be a
set $\SS=\set{R_1\dd R_n}$ of subregions such that $R=\cup_{i=1}^n R_i$
and the interiors of $R_i$'s are pairwise disjoint.
If each $R_i$ is a box, we call $\SS$ a \dt{box subdivision}.
The box subdivision is a \dt{box complex}
if for any two adjacent boxes $B,B'\in\SS$,
their intersection $\partial(B)\cap \partial(B')$ is side of either $B$ or $B'$.
Clearly, the set $\SS$ of leaf boxes of $\TT$ forms a box complex of $B_0$.
But in this paper, we need to consider a more general
subdivision of $B_0$ that is obtained as the leaf boxes of a finite number of
subdivision trees.
A \dt{segment} of a box complex $\SS$ is the side of a box of $\SS$ that does
not properly contain the side of an adjacent box.  Therefore every side of
a box of $\SS$ is a finite union of segments.
We say the box complex $\SS$ is \dt{balanced} if every side is either
a segment or the union of two segments. A segment is called \dt{bichromatic}
w.r.t. a curve $S$ if $S$ has different signs on the endpoints of the segment;
otherwise call it \dt{monochromatic}.

Although $(S,T)$ is simple, we need to
consider degeneracies \dt{induced} by a subdivision $\SS$:
\alt{
we say $(S,T)$ is \dt{$\SS$-regular} if $S\cup T$ does not intersect any corner
of a box in $\SS$.  This can be effectively achieved by an infinitesimal
perturbation of $S$ and $T$ using a trick in \cite{plantinga-vegter:isotopic:04}:
when we evaluate the sign of $f$ at a box corner, we simply regard a $0$ sign to be $+1$.
}{
	\bitem
	\item[(R1)]
	The union $S\cup T$ intersects a corner of a subdivision box of $\SS$.
	\item[(R2)]
	The union $S\cup T$ has tangential intersection with
	an edge of a subdivision box.
	\item[(R3)]
	The intersection $S\cap T$ intersect an edge of a subdivision box.
	\eitem
These conditions are illustrated in \refFig{normalized}.
{\em
Henceforth, we may assume that that our original input arrangement
$(S,T)$ is $\SS$-regular (for any $\SS$ of interest).}
It is clear that for arbitrary $(S,T)$, we can perturb $(S,T)$ into
a $\SS$-regular arrangement (for any desired $\SS$) by
applying an infinitesimal perturbation to $(S,T)$.
In practice, if $S=f^{-1}(0)$, when we need to evaluate the sign of $f$ at
a corner of a box of $\SS$, we simply regard a sign to be $+1$ when it is actually $0$.
This simple trick was used in \cite{plantinga-vegter:isotopic:04}.
}

\myPara{Normalization.}

Consider an isotopy of the arrangement $(S,T)$ into another arrangement $(S',T')$.
Let us write $(S,T)_t$ for the arrangement at time $t\in [0,1]$
during this transformation.  Thus $(S,T)_0=(S,T)$ and $(S,T)_1 = (S',T')$.
The isotopy is said to \dt{$\SS$-regular} provided, for all $t\in[0,1]$,
$(S,T)_t$ is $\SS$-regular.
We say that $(S,T)$ is \dt{$\SS$-normalized} if:
\begin{mini_itemize}
\item [(N0)]
$(S,T)$ is $\SS$-regular.
\item[(N1)]
Each subdivision box $B$ of $\SS$ contains at most one point of  $S\cap T$.
\item[(N2)]
Let $X\in \set{S,T}$.  Then $X$ intersects each segment of $\SS$ at most once
\end{mini_itemize}

Call $(S',T')$ a \dt{$\SS$-normalization} of $(S,T)$
if there exists a $\SS$-regular isotopy from $(S,T)$ to $(S',T')$ such that
$(S',T')$ is $\SS$-normalized.
Our algorithm will construct an $\SS$-normalization $(S',T')$ of $(S,T)$.

\myPara{Box Predicates.}
We will use a variety of box predicates.  These predicates will determine
the subdivision process.  Typically, we will keep subdividing boxes until
some Boolean combination of some box predicates hold.

Let $h:\RR^2\to\RR$ be any real function.
Recall (Appendix A) that we assume an interval formulation of $h$
denoted $\intbox h:\intbox\RR^2\to\intbox \RR$ where $\intbox \RR$ denotes
the set of closed intervals and $\intbox\RR^2$ can be viewed as the set of boxes.
We introduce a pair of box predicates denoted $C_0^h$ and $C_1^h$, defined as
	\beql{c01}
	\grouping{
		C_0^h(B) &\equiv& 0\nin \intbox h(B),\\
		C_1^h(B) &\equiv& 0\nin (\intbox h_x(B))^2 + (\intbox h_y(B))^2.}
	\eeql
Note that $C_1^h$ as taken from Plantinga-Vegter, 
where the interval operation $I^2$ is defined as $\set{xy: x, y\in I}$ and
not $\set{x^2: x\in I}$.
An alternative to $C_1^h$ would be the weaker $C_{xy}^h$ predicate from Lin-Yap \cite{lin-yap:cxy:11},
but the corresponding algorithm would would be more involved.  So for now,
we focus on the $C_1^h$ predicate.  We classify boxes using these predicates:
\begin{mini_itemize}
	\item Box $B$ is \dt{$h$-excluded} if it satisfies $C_0^h(B)$.
	\item Box $B$ is \dt{$h$-included} if it fails $C_0^h(B)$ but satisfies $C_1^h(B)$.
	\item Box $B$ is \dt{resolved} if it satisfies the predicate
		\beql{resolved}
		(C_0^f \lor C_1^f) \land (C_0^g \lor C_1^g).
		\eeql
	\item Box $B$ is \dt{excluded} if it satisfies
		$C_0^f \land C_0^g$.
		Note that excluded boxes are resolved.
	\item Box $B$ is a \dt{candidate} if it is resolved but not excluded. 
	\item Candidate boxes can be further classified into three subtypes:
		\dt{$f$-candidates} are those that are $f$-included but $g$-excluded,
		\dt{$g$-candidates} is similarly defined, and
		\dt{$fg$-candidates} are those that are $f$- and $g$-included.  
\end{mini_itemize}

\myPara{Root Boxes.}
We define a \dt{root box} to be any box $B$ where $B\cap S\cap T$ has
exactly one point.
We next consider two predicates that will allow us to detect root boxes.
One is the \dt{Jacobian condition}, 
	$$\JC(B)\equiv 0\notin \det(\intbox J_F(B))$$
where $\intbox J_F(B)$ is the Jacobian of $F=(f,g)$ evaluated on $B$.
If $\JC(B)$ holds, then $B$ has at most one root of $f=g=0$,
The other is the \dt{Moore-Kioustelidis condition} $\MK(B)$ \cite{mk:test:80}
which can be viewed as a preconditioned form of the famous
Miranda Test \cite{kulpa:poincare-miranda:97};
for other existence tests based on interval arithmetic
see \cite{frommer-lang:miranda-type-tests:05}.
If $\MK(B)$ holds, then $B$ has at least one root of $f=g=0$.
We provide the details for this predicate in Appendix B; see \refeq{mktest}.
Therefore, when $\JC(B)$ and $\MK(B)$ holds, we know that $B$ is a root box.
The use of Miranda's test
combined with the Jacobian condition has been used earlier to isolate the common roots 
\cite{mantzaflaris+2:continued-frac:11}.
What is new in this paper is its application to the simple curve arrangement problem.

\subsection{Graph Representation}

Our algorithm will produce a graph $G=(V,E)$ where
vertices $v\in V$ are points in $\RR^2$ and edges
are line segments connecting pairs of vertices.
Moreover, each edge $E$ will be labeled as an $S$-edge or a $T$-edge.
The union of these edges will provide a polygonal $\eps$-approximation of $(S,T)$.
We now give an overview of the issues and solution.

First, we describe how the vertices of $V$ are introduced.

\begin{mini_itemize}
\item[(V0)]
We introduce a vertex in the center of a root box $B$.

\item[(V1)]
We evaluate $f,g$ at the endpoints of segments of $B$.  
If $h\in\set{f,g}$ is bichromatic on a segment of $B$, then we must
introduce an \dt{$h$-vertex} somewhere in the segment. 
In a balanced subdivision, an $\SS$-normalized pair $(S',T')$ of curves has
at most two $h$-vertices on an edge of a box $B$.

\item[(V2)]
Introducing vertices on the edges of a box $B$ is straightforward if $B$ is an $f$-candidate
or a $g$-candidate.   When $B$ is a $fg$-candidate, we may have an edge $e$
containing both a $f$-vertex and a $g$-vertex.  In the next section we will  show
how to find the relative order of these two vertices.
 
\end{mini_itemize}

Next we discuss how to introduce the edges $E$, which are line segments completely contained in a box.

\begin{mini_itemize}
\item
If $B$ is a root box, we just connect the vertex at its midpoint $\cen(B)$
to each of the vertices on the edges of $B$.
There will be exactly two $f$-vertices and two $g$-vertices.
\item
If $B$ is a $f$-candidate or $g$-candidate, then the connection is trivial
in the regular case.  In the balanced case, the rules from the 
previous work of Plantinga-Vegter \cite{plantinga-vegter:isotopic:04}
assures us of the correct connection.
\item
If $B$ is a $fg$-candidate, but not a root box,
we know that the $f$-segment and $g$-segment will not intersect.
Some $fg$-candidates need global information to resolve them: when there are
two edges where each edge contains both an $f$- and a $g$-vertex.
Their relative order must be determined globally from root boxes or from
boxes where their relative order is known.  We will show how to propagate this information in \refSec{algo}.
  
\end{mini_itemize}

\subsection{Curve Arrangement in Root Boxes}

Suppose $(S',T')$ is the normalization of $(S,T)$ relative to
the box $B$, i.e., $(S',T')$ is an isotopic transformation
of $(S,T)$ which respects the four corners of $B$.
We now determine the isotopy type of $(S',T')$ in a root box $B$.
The possible combinatorial types fall under one of the $8$
patterns as shown in \refFig{patterns}. We put them in three groups (I, II, III) for our analysis.

\figchoose{

	    \begin{figure}[htb]
	    \begin{center}
		\scalebox{0.5}{
	    	  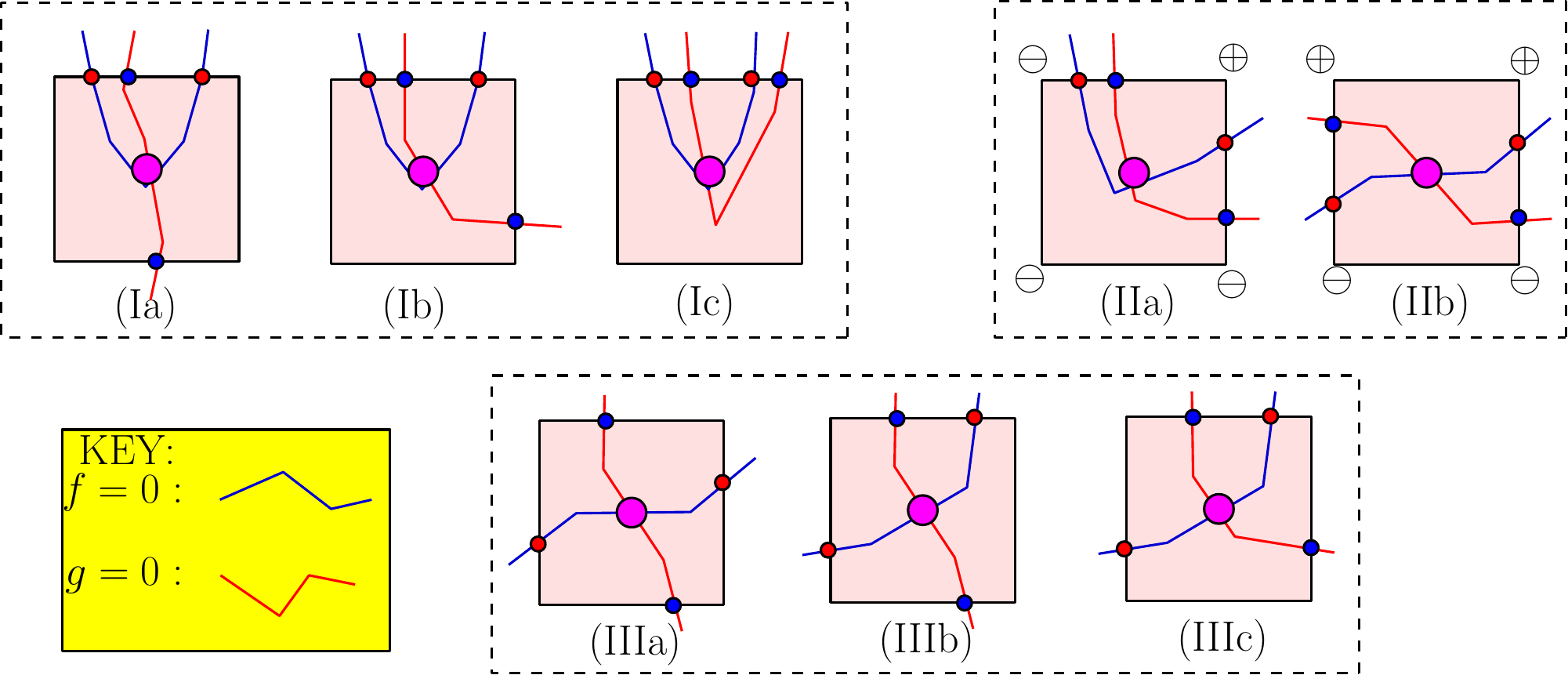}
	    \caption{Local intersection patterns of the normalized curves $(S',T')$}
	    \label{fig:patterns}
	    \end{center}
	    \end{figure}
	}{
\vfigpdf{Local intersection patterns of the normalized curves $(S',T')$}{patterns}{0.5}}

Following the standard Marching Cube technique, we evaluate the sign of the functions
$f,g$ at the four corners of $B$.
If $f$ has different signs at the endpoints of an edge $e$ of $B$, then we must
introduce an \dt{$f$-vertex} somewhere in the interior of $e$.   
Our normalization assumptions imply that there are either zero or
two $f$-vertices on the boundary of $B$. We treat $g$ similarly. 
Our aim is to connect the two $f$-vertices, the two $g$-vertices, and
a point in the center of the box which represents the common root with
line segments such that the graph $G$ obtained 
is an isotopic approximation of $(S'\cap B_0, T'\cap B_0)$.
There is a subtlety: the method exploits ``local non-isotopy''
\cite{plantinga-vegter:isotopic:04,lin-yap:cxy:11}, meaning that we do not
guarantee that $S\cap B$ is isotopic to the segment introduced to connect two $f$-vertices.
However, the graph $G$ will be locally isotopic to the normalized curves $(S',T')$, i.e., $G\cap B$
is isotopic to $(S'\cap B, T'\cap B)$ in each subdivision box $B$.

The issue before us is the relative placements of an $f$-vertex and $g$-vertex
in case they both occur in $e$; e.g., the patterns in group II in \refFig{patterns}.
The main result of this section is the following.

\bthml{relorder}
Let $B$ be a root box that satisfies $\MK(B)$.
Then the signs of $f$ and $g$ at each of the four corners of $B$
determine the combinatorial type of the normalized
curves $S',T'$ in $B$.  Moreover, these combinatorial types
fall under one of the five types in Groups II and III in \refFig{patterns}.
\ethml
The main idea of the proof is that if $\MK(B)$ holds for a box $B$ then  there
exists an edge $e$ of $B$ such that either $f(e) > c g(e)$, or $g(e) > cf(e)$, for some $c > 0$.
Given such an $e$, we can find the relative order of the $f$-vertex and $g$-vertex on $e$.
See Appendix C for details of the proof.
\ignore{{\em if a subdivision box $B$ satisfies $\JC(B)$ and $\MK(B)$, then
by computing the signs of $f,g$ at each of the four corners of $B$,
we can determine the combinatorial pattern of the normalized
curves $S',T'$ in $B$.  Moreover, these combinatorial pattern
fall under one of the five types in Groups II and III.
}}

\subsection{Geometry of Extended Root Boxes}

By an \dt{aligned box} we mean one that can be obtained
as a node of a subdivision tree rooted at the region-of-interest (ROI) $B_0$; 
otherwise, it is said to be \dt{non-aligned}.
For instance, in \refFig{aligned-new}(a), let the box with corners
$p,q,r,s$ be $B_0$. Then the figure shows the four children of
$B_0$, which are aligned, as well as the non-aligned box $(1/2)B_0$ whose corners are $p',q',r',s'$.
Note that $(1/2)B_0$ can be obtained as the union of aligned boxes.  We are interested in non-aligned
boxes that can be obtained as a finite union of aligned boxes.  In the simplest case
of non-alignment, a box $B$ is said to be \dt{half-aligned} if it is
equal to the union of congruent aligned boxes of size $w(B)/2$.
Thus if $B$ is aligned then both $(1/2)B$ and $2B$ are half-aligned.

\figchoose{
\myfigfloat[3in]{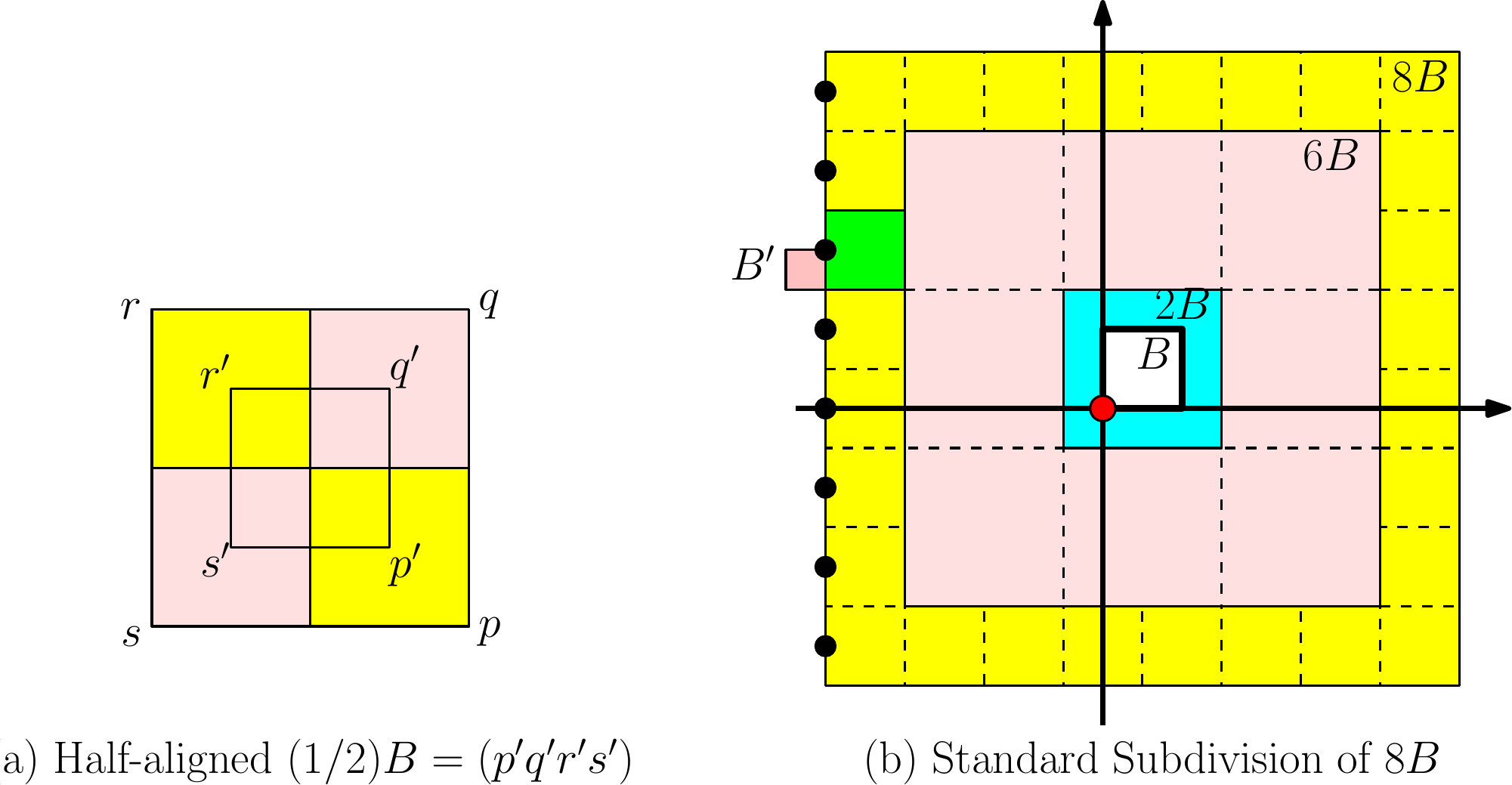}{(a) $B=(pqrs)$ is aligned,  (b) $2B$ is a root box
	}{aligned-new}{0.3}
}{
\vfigpdf{(a) $B=(pqrs)$ is aligned,  (b) $2B$ is a root box. }{aligned-new}{0.5}}

In most subdivision algorithms, it is enough to work with aligned boxes.
But to treat root boxes, we see an essential need to work with non-aligned boxes.
The reason is that if we apply the Moore-Kioustelidis predicate
to aligned boxes, non-termination may occur when a root of $F$ lies on 
the boundary of an aligned boxes.  But such roots can be detected
in the interior of non-aligned boxes.
This issue is often ignored in the literature, but it needs to be
properly treated in exact algorithms.  Some discussions may be found
in Stahl \cite{stahl:thesis:95} and Kamath \cite{kamath:thesis}; in the univariate
case, a solution is suggested by Rote \cite{rote:subdivision:08} for splines.

Therefore, given an aligned box $B$, we provide a procedure to detect if $2B$ is a root box.
We consider the nested sequence of boxes
	$B\ibp 2B \ibp 6B\ibp 8B$
as illustrated in \reffig{aligned-new}(b).
Our goal is to detect $2B$ as a root box, but because of alignment issues,
we must also treat the larger box $8B$ which is called the \dt{extended root box}
corresponding to $B$.

We construct the following \dt{standard subdivision} of $8B$, denoted $\std(B)$, into sub-boxes:
\begin{mini_itemize}
  	\item
          Subdivide $6B$ into $9$ boxes, each congruent to $2B$ (indeed, $2B$ is one of
	these $9$ boxes).
	\item
	The annular region $8B\setminus 6B$ is partitioned into $28$ boxes, each congruent to $B$.
	These are called the \dt{ring boxes}.
\end{mini_itemize}
See \refFig{aligned-new}(b) for illustration. Note that $\std(B)$ is balanced.   
None of the subdivision boxes are aligned, but the ring boxes are half-aligned.

\myPara{Conforming Subdivisions.}
Let $\Pi$ be a subdivision of a region $R$.
A box $B'\in \Pi$ is a \dt{boundary box} of the subdivision
if $\partial B'$ intersects $\partial R$. 
In the following definitions, we fix a region $R_0\ib B_0$ and
fix a box $B$ such that $8B\ib R_0$. Also let $k\ge 1$ be an integer. 

A subdivision $\Pi_0$ for $R_0\setminus 8B$ is called \dt{externally $k$-conforming for $B$} if 
it has three properties:
$\Pi_0$ is balanced, the union $\Pi_0 \cup \set{8B}$ is a box complex, and for each
box $B'\in \Pi_0$,  if $B'$ is adjacent to $8B$	then $w(B')=w(B)/2^k$. 
A subdivision $\Pi_1$ of $8B$ is called \dt{internally $k$-conforming for $B$} if
$\Pi_1$ is balanced, and
for every boundary box $B'$ of $\Pi_1$, $w(B')=w(B)/2^{k-1}$.
Note for instance that if $\Pi_1$ is the standard subdivision of $8B$, then
it is internally $1$-conforming for $B$.
Below we show how to achieve subdivisions of $8B$ that
is internally $k$-conforming for $B$ for $k\ge 2$. The following is immediate:
{\em
If $\Pi_0$ is externally $k$-conforming for $B$,
and $\Pi_1$ is internally $k$-conforming for $B$,
then their union $\Pi_0\cup\Pi_1$ is a balanced subdivision of $R_0$.
} 
Note that if $k > 1$ then getting a balanced subdivision of $\Pi_0 \cup \Pi_1$
may cause the edges of a root box $2B$ to split into two segments (but not more);
see \refFig{conform}. 
This can be handled by a case analysis similar to \refThm{relorder} based on \refLem{fgorder}.
An alternative approach is to replace $8B$ by $10B$
which would have an extra ring of boxes congruent to $B$.
In this case, we can handle any $k>1$ by subdividing this outermost ring,
but without affecting the standard subdivision of $8B$.
This gives a simple and effective solution.

\myPara{Strong Root Isolation.}
Suppose $2B$ is a root box.  We say $2B$ is \dt{strongly isolated} if the following
conditions hold
\begin{mini_itemize}
  	\item (P1) The following four predicates hold:
		$C_1^f(8B), C_1^g(8B), \JC(6B), \MK(2B)$.
	\item (P2) $F=(f,g)$ has no roots in the annulus $8B\setminus 2B$.
\end{mini_itemize}
The predicates in (P1) ensures that $2B$ is a root box.
It is not hard to see that if $2B$ contains a root of $F$ and is sufficiently small,
then properties (P1) and (P2) will hold.
The reason for $\MK(2B)$ (not just $\MK(B)$ is to ensure that we test the 
Moore-Kioustelidis predicate on overlapping boxes, so that roots
on the boundary of an aligned box $B$ will appear in the interior of
$2B$.
The reason for $\JC(6B)$ instead of $\JC(2B)$ is that there can be two boxes 
$2B$ and $2B'$ such that both of them satisfy MK-test and they overlap.
The test $\JC(6B)$  ensures that if there are two such boxes then they correspond to the same
root, and so discard one of them.


\myPara{Root Refinement:}
Let $B$ be an aligned box from the subdivision queue such that $2B$ is a root box.
We give a subroutine to refine such a root box $2B$.
It it important that in our refinement method 
all the sub-boxes remain dyadic boxes, assuming the input boxes are dyadic.
The idea is to cover $2B$ with a covering of aligned boxes, which must be of size $w(B)/2$,
and check whether MK-test holds for the doubling of any of these 16 boxes. If not, then 
subdivide these boxes and continue recursively with the $fg$-candidates. See Appendix A for more details.

\ignore{
\myPara{Conformal Subdivision}
Suppose $8B_1\dd 8B_r$ is a collection of pairwise disjoint extended root boxes,
and moreover, there are no roots in their complement,
	$B_0 \setminus \left(\cup_{i=1}^r 8B_i\right)$.
We first compute a balanced subdivision for this in the usual way, until every leaf box is placed in one of the
queues $Q_0, Q_f, Q_g, Q_{fg}$.

The problem before us is to compute a balanced subdivision of the extended root boxes $8B$'s
and of their complement in $B_0$.
The region $B_0\setminus \bigcup_{i=1}^r 8B_i$ can be represented by a subdivision tree $\TT$.
rooted at $B_0$ (see \cite{lin-yap:cxy:11}).
We will say the subdivision $\Pi$ is \dt{conformal} if for any box $B$ in $\Pi$, if
$B$ is adjacent to $8B_i$, then $B$ is congruent to $(1/2)B_i$. 
The first goal of our algorithm is compute a complete set of root boxes,
and its associated conformal subdivision tree $\TT$.
}

\section{Algorithm for Curve Arrangement}\label{sec:algo}

\ignore{
In numerical algorithms, a key idea is that of ``filters''.   
In trying to ensure that a box $B$ satisfies a certain property $C(B)$,
we may first apply some relatively cheap predicate $C'(B)$ which has
the property that if $C'(B)$ holds then $C(B)$ does not hold.
Thus, if $C'(B)$ holds, there is no need to evaluate $C(B)$.
We call $C'$ a 	``exclusion filter'' for $B$.
There is a similar concept of ``inclusion filter''.
Note that if $C'(B)$ fails, we have no conclusion whether $C(B)$ holds or not.  
This filter idea will motivate the order in which we apply various
predicates in our algorithm.
}

Our overall algorithm begins with the (trivial) subdivision tree $\TT$ rooted at the ROI $B_0$ but
with no other nodes.   The algorithm amounts to
repeatedly expansion of the candidate leafs in $\TT$ until a variety
of global properties hold. We  given an overview of the algorithm in a sequence of 9 \dt{stages};
see Appendix C.

\myPara{Stage I: Resolution Subdivision}
The high level description of this stage is easy:
keep expanding any leaf $B$ of $\TT$ that is not resolved
(see \refeq{resolved}).
Recall that resolved boxes are either excluded or candidates.
As each box is resolved, it is placed in one of the following four queues:
$Q_0$ for excluded boxes, $Q_f$ for $f$-candidates, $Q_g$ for $g$-candidates, and $Q_{fg}$ for $fg$-candidates
Besides these four global queues, we also use these additional queues:
	$Q_{\JC}, Q_{\MK}, Q_{\Root}$
corresponding roughly to boxes that satisfies the $\JC$ and $\MK$ predicates,
or are found to be root boxes. {\em The boxes in all the queues are always aligned boxes}.

\myPara{Stage II: Jacobian Stage.}
Remove a box $B$ from $Q_{fg}$ and do the following: If $\JC(6B)$ holds then put $B$ into $Q_\JC$,
otherwise, subdivide $B$ and distribute the children into $Q_0, Q_f, Q_g, Q_{fg}$.

\myPara{Stage III: MK Stage.}
        For every box $B \in Q_\JC$ we subdivide it until either we find a sub-box $B'$ such that
        $\MK(2B')$ holds, or we have identified all sub-boxes as one of $Q_0, Q_f, Q_g, Q_{fg}$.

\myPara{Stage IV: Strong Root Isolation Stage}
	We assume that $Q_\MK$ is a priority queue, where boxes are popped
	starting from the largest size. For each such box $B$ check whether $8B$ is disjoint
        from $8B'$, for all its neighbors $B'$; if not then replace $B$ with RefineRoot$(B)$.
       We now have obtained a queue $Q_\Root$ containing root boxes for all the roots in ROI. 
        The next step is to externally conform $\std(B)$ with the rest of the subdivision tree $\TT$.

\myPara{Stage V: Pruning $\TT$}
        In this stage we will turn OFF some leaf boxes in $On(\TT)$ depending on how they
        interact with the extended root boxes $8B$. The aim is to ``blackout'' the $8B$ regions from ROI,
        and ensure that the boxes abutting it are all aligned boxes. Let $B'$ be the great-grandparent of $B$
        in $\TT$. Then we get the list of leaf boxes that cover the interior of $B'$ and another list of boxes
        that are its neighbors. For each box $B_\tmp$  in these lists, we turn it OFF if it is contained in $8B$;
        if it overlaps $8B$ then we subdivided it and proceed with its children. Let $\TT'$ be the resulting
        subdivision       tree.

\myPara{Stage VI: Balancing and Externally Conforming}
        Recall the standard balancing procedure for a subdivision $\TT$ of a region $B_0$ from the appendix. 
        We will construct a balanced and externally conformal subdivision of $B_0 \setminus \cup_i 8B_i$, where
        $8B_i$'s are pairwise disjoint extended root boxes. For each box $8B_i$, we add a conceptual
        box to $\TT'$, with depth either one more than its smallest neighbor, or if all the neighbors of $8B$
        are larger than $w(B)$        then one more than the depth of $B$ in $\TT$. Call the standard
        balancing procedure on the modified $\TT'$. By \refLem{balance}, we will get the desired subdivision;
        after balancing the boxes bordering $8B$ will all be of the same size, namely $w(B)/2^k$, for some $k \ge 1$.

\myPara{Stage VII: Internally Conforming Extended Root Boxes}
        Consider any extended root box $8B$ and its standard subdivision $\std(B)$. Given a $k > 1$ from
        the previous stage, we want to balance the interior and the exterior of $\std(B)$. Note that since $k > 1$
        the boxes  on the exterior are always smaller than all the boxes in $\std(B)$. To get a balanced conformal
        subdivision  of $\std(B)$, we initialize a priority queue $Q$ with all the boxes on the exterior of $8B$
        (all of them are of the same size) and the 37 boxes in $\std(B)$. Then we initiate the standard balancing
        procedure on $Q$.  See \refFig{conform}(c)
        for an illustration of this procedure; the box $B'$ has width $w(B)/8$.
        \ignore{
          The balancing property, \refLem{balance}, ensures that after we have balanced 
        the 28 ring boxes of $\std(B)$ to make them  conformal with the exterior, 
        the boxes that border $6B$ are of size $\ge w(B)/2$. These
        boxes thus may not be balanced w.r.t. the eight $2B$ boundary boxes in the subdivision of $6B$. 
        However, a single subdivision of each of these $2B$ boxes will make them balanced w.r.t. their inner
        neighbor, which is the root box $2B$, and w.r.t. their external neighbors.}
        We do this balancing step         for each of the extended root boxes $8B$. The union of these subdivisions with the balanced subdivision of
        $B_0 \setminus \cup_i 8B_i$ gives us a balanced subdivision of $B_0$, our ROI.
        \ignore{At the end of
        this stage, we additionally ensure that each box in the subdivision of $8B$ is put in one of
        the queues $Q_0$, $Q_f$, $Q_g$ and $Q_{fg}$; note only $B$ goes to $Q_\Root$.}
        
\myPara{Stage VIII: PV-Construction}
        For each box in $Q_f$, connect its two $f$-vertices with a line segment; do the same for boxes
        in $Q_g$. For each box in $Q_\Root$  place a vertex at its center and connect
        the two $f$-vertices and the two $g$-vertices with this vertex according to the cases shown in Groups II and III.
        of \refFig{patterns}.
        At the end of this stage, the only queue that remains unprocessed is $Q_{fg}$. The next stage
         resolves these boxes.

\myPara{Stage IX: Resolving Ambiguous $fg$-candidates}
        We call an $fg$-candidate box \dt{ambiguous} if they have the same set of bichromatic segments;
        otherwise, call the box \dt{unambiguous}. By definition, boxes where 
        $f$ and $g$ do not share a bichromatic segment are unambiguous. However, some
        ambiguous boxes can be made unambiguous locally. 
        From \refThm{relorder} we know that ambiguous root boxes can be made unambiguous.
        Also, boxes where the two shared bichromatic segments are on adjacent edges can be
        made unambiguous by repeated subdivisions of the edges until we reach a segment in
        one of the edges that is bichromatic for one curve and monochromatic for the other; this will
        happen along one of the edges since both $C_1^f$ and $C_1^g$ hold.
        A similar approach works to resolve ambiguous boxes that share an edge with $B_0$ and a
        common bichromatic segment is on this edge, because by assumption boundary of $B_0$ 
        does not contain a root of $f,g$.
       From these unambiguous boxes, we propagate the ordering of 
        the $f$-vertex and $g$-vertex on the shared edge to their ambiguous neighbors.
\figchoose{
\FigEPS{conform}{0.30}{An internally conformal subdivision of $\std(B)$}}
{\vfigpdf{An internally conformal subdivision of $\std(B)$.}{conform}{0.35}}

\myPara{Correctness of Algorithm}
We must prove that our graph $G=(V,E)$ is isotopic to the arrangement $(S,T)$ 
in box $B_0$.   Suppose there are $k$ roots, $|S\cap T|=k$.  
{\em Our correctness requires that none of these roots lie in $\partial B_0$.}
Our algorithm produces the following data:  we have ``well isolated'' the roots
in this sense: we have found $k$ aligned boxes,
$B_1\dd B_k$ such that $2B_i$ is a root box,
$8B_i\ib B_0$, and the interiors of the $8B_i$'s are pairwise disjoint.
Next, we have constructed subdivisions,
	$$\SS_0, \SS_1\dd \SS_k$$
where $\SS_i$ is a subdivision of $8B_i$ ($i=1\dd k$) and $\SS_0$ is
a subdivision of $B_0\setminus \cup_{i=1}^k 8B_i$.
Moreover, the union of all these subdivisions, denoted $\SS^*$,
constitutes a balanced box complex of $B_0$.

\bthml{main}
The PSLG $G$ computed by the algorithm is a $\SS^*$-normalization of the curves $(S,T)$.
\ethml

We sketch the arguments here: let $(S',T')$ be a $\SS^*$-normalization of $(S,T)$.
The graph $G$ will be obtained as the union of $G_B$ for all $B\in \SS^*$,
where each $G_B$ is a PSLG contained in box $B$.
We know from Theorem 1 how to construct a PSLG $G_B\ib B$ that is isotopic to $(S',T')$
in each root box $B$.   We know from Plantinga-Vegter how to construct 
PSLG $G_B^S$ that are isotopic to $S'$ in each non-root box $B$.
Similarly we have $G_B^T$.   But we need to form their "union", which is
the PSLG $G_B$ that is isotopic to $(S',T')$ in $B$.
For this purpose, we need to know the relative
ordering of the $f$-vertex and $g$-vertex on each segment of $B$
that is bichromatic for both curves.   This information is resolved by Stage IX
of our construction.



\section{Final Remarks}

We have presented a complete numerical algorithm for the isotopic
arrangement of two simple curves.   The underlying paradigm is
Domain Subdivision, coupled with box predicates and effective
forms of the Miranda Test.  Moreover, we crucially exploit the previous
isotopic approximation algorithms of Plantinga-Vegter \cite{plantinga-vegter:isotopic:04}
for a single curve.

The algorithm is very implementable: despite the many stages, each stage
involves iteration using well-known data structures.   A full implementation
and comparisons with other methods is planned; we have currently 
implemented the root isolation part.     

The extension of this work to the simple arrangement of multiple curves is
of great interest.  Many of the techniques we have developed for 2 curves will
obviously extend.  One possible way to use our work for multiple curves
is as follows: first compute the root boxes $2B_i$ of all the pairwise intersections,
and make them ``well isolated'' in the sense that $8B_i$ boxes are pairwise disjoint,
as before.   Then we
compute a balanced, conforming subdivision $\SS_0$ of complement of the
union of these $8B$ boxes.  Moreover, we need to resolve ambiguities, i.e.,
relative ordering of curves on a common segment.  Some of this can be resolved
by propagation, but there will be need for recursive subdivision in general.
In the full paper, we will provide such a description.

A general open problem is to prove polynomial complexity bounds
for such subdivision algorithms.   As a first step, we would like to
prove that the root isolation part is polynomial-time.  This would be
a generalization of our recent work on continuous amortization
for real and complex roots \cite{sagraloff-yap:ceval:11}.



\newpage
\begin{appendices}
\section{Basic Concepts}
We fix the terminology for well-known concepts in
boxes, interval arithmetic and subdivision trees. We define these concepts
in $d$-dimensions. Of course, the algorithms in this paper work in $d=2$.

\myPara{Boxes.}
Let $\intbox\RR$ denote the set of closed intervals.  We may identify
$\RR$ with degenerate intervals $[a,a]\in\intbox\RR$.
Also $\intbox\RR^d$ is the $d$-fold Cartesian product of $\intbox\RR$.
Elements of $\intbox\RR^d$ are called \dt{$d$-boxes}.
The \dt{width} of $B$ is $(w(I_1)\dd w(I_d))$ where
the width of an interval $I=[a,b]$ is $w(I)=b-a$.
the same (resp., differ by at most $1$). 
If $B, B'$ are two boxes in $\intbox \RR^d$, we say they are \dt{$k$-neighbors} if
$B\cap B'$ has dimension $k$.  So $k\in \set{-1,0,1,d-1}$, where the empty set
has dimension $-1$.   
We say $B$ and $B'$ are \dt{adjacent} if they are $(d-1)$-neighbors.
Each box has $2^d$ \dt{sides} (sometimes called \dt{edges}) and $2^d$ \dt{corners}.
The boundary of a box $B$ is denoted $\partial B$.

\myPara{Box Functions.}
Interval arithmetic \cite{moore:bk} is central to our computational toolkit.
If $f:\RR^d\to\RR$ is a real function, then we call a function of
the form $\intbox f:\intbox\RR^d\to\intbox\RR$ an \dt{inclusion function}
for $f$ if for all $B\in\intbox\RR^d$, $\intbox f(B)$ contains $f(B)=\set{f(p): p\in B}$.
Call $\intbox f$ a \dt{box function} for $f$ if it is an inclusion function for $f$
and for all $\set{B_i\in \intbox\RR^d : i\in\NN}$, if $B_i$ converges monotonically to a point $p\in\RR$
then $\intbox f(B_i)$ converges monotonically to $f(p)$.
Note that box functions are easy to construct for polynomials and common real functions.

\myPara{Subdivision Trees.}
Our fundamental data structure is a quad-tree
or \dt{subdivision tree} $\TT$: the nodes of $\TT$ are boxes in $\intbox\RR^d$,
and each internal node $B$ has $2^d$
children which are congruent sub-boxes, with pairwise disjoint interiors,
and whose union is $B$.  In order to use $\TT$ to represent regions
of complex geometry, we assume that
each leaf of $T$ is (arbitrarily) either turned ON or turned OFF.
The union of all the ON-leaves is denoted $R(\TT)$,
called the \dt{region-of-interest} (ROI). 
Let $On(\TT)$ denote the set of ON-leaves of $\TT$.
We call $On(\TT)$ a \dt{subdivision} of $R(\TT)$.
In general, a \dt{subdivision} of a set $X\ib\RR^d$ is
a collection $C$ of sets in $\RR^d$ such that $\cup C = X$ and
the relative interior of the sets in $C$ are pairwise disjoint.
%
One of the basic operations on subdivision trees is to take an ON-leaf $B$ of $\TT$
and to ``expand it'', i.e., to split $B$ into $2^d$ congruent sub-boxes
and attach them as children of $B$.   Thus $B$ becomes an internal node
and its children become leaves of the expanded $\TT$.   By definition,
the children of $B$ remain ON-leaves.  Thus the ROI is not affected by expansion.

A \dt{segment} of $\TT$ is a line segment of the form $B\cap B'$ where
$B,B'$ are adjacent boxes in $\TT$.   Note that a segment is always an edge of some box,
but some box edges are not segments.  In general, an edge is
subdivided into a finite number of segments.

The boxes of a subdivision tree are assumed to be non-degenerate, i.e.,
they are $d$-dimensional.  
In our algorithms, certain ON-leaves are called ``candidates box''.
Unless otherwise noted, we could assume every ON-leaf is a candidate box.
We then say $\TT$ is
\ignore{\footnote{
	Note that in our previous work
	(\cite{plantinga-vegter:isotopic:04,lin-yap:cxy:11}),
	uniform subdivisions were called ``regular subdivisions''.
	The current usage of regular/uniform seems better. 
}}
\dt{balanced} if, for any two candidate boxes, if they are adjacent then
their depths differ by at most one.

\textbf{Traversing neighbors in a subdivision of ROI:} Given a subdivision tree $\TT$ 
partitioning the ROI, a crucial sub-procedure required by the algorithm is the ability 
to get the neighbors of a leaf-box in $\TT$.  One way to achieve this is to associate two pointers with 
every edge of a leaf box of $\TT$, namely the pointers that point to the extreme
neighbors along the edge (there may be only one such neighbor, in which the two pointers
point to the same box). Thus we associate 8 pointers with every leaf-box. 
We will often say the ``eight neighbors'' of a box
to refer to the boxes pointed by these eight pointers, where we count the boxes with multiplicity.
We can list all the neighbors of a leaf-box $B$ in $\TT$ using these eight pointers.

\textbf{ Standard Balancing Procedure:}
        \progb{
          Let $Q_\tmp$ be a priority queue of all the leaves in $\TT$;\\
	  the deeper the level the higher the priority.\\
          While $Q_\tmp$ is non-empty do\\
          \> $B \ass Q_\tmp.pop()$.\\
          \> For each neighbor $B_\tmp$ of $B$ do\\
          \>\> If  $B_\tmp$ is not balanced w.r.t. $B$ subdivide\\
	  \>\> $B_\tmp$ and add its children to $Q_\tmp$.
       }
There can be at most two neighbors of $B$ that need to be subdivided, because $B$ shares two edges
with its siblings and so the boxes neighboring $B$ along those edges are balanced w.r.t. $B$;
the unbalanced boxes can occur on the remaining two edges. Moreover, for any neighbor $B_\tmp$
that is subdivided only one of its children neighbors $B$. Balancing also has the following 
nice property, which intuitively says that the boxes produced in the ensuing subdivision cannot
all be very small.

\bleml{balance}
Suppose we are balancing a box $B$, and let $B'$ be its violating larger neighbor.
Let $e$ be the edge of $B'$ shared with $B$ and $e'$ be the  opposite edge. 
Then the subdivision of $B'$ caused by $B$ while balancing 
will split the edge $e'$ only once.
\eleml
In the subdivision tree of $B'$, the two children that share $e'$ are in a different
subdivision tree compared to the child of $B'$ that is adjacent to $B$ and shares $e$; see \refFig{balance}. 
Balancing produces a subdivision tree of $B'$ that has only one path, with leaves hanging from it,
that ends in a box whose size is double the size of $B$. The number of leaves in this tree are
$3 \cdot (\log w(B') - \log w(B) - 1)$.
\figchoose{
\FigEPS{balance}{0.30}{A subdivision caused balancing}}
{\vfigpdf{A subdivision caused balancing.}{balance}{0.35}}

\newpage
\section{The Moore-Kioustelidis Test for Roots}

Although our paper is focused on arrangement of curves, we shall temporarily
consider a more general setting of a continuous function $F:\RR^n\to\RR^n$ in $n$-space.
Let the coordinate functions of $F$ be denoted $(f_1\dd f_n)$.
If $B=\prod_{i=1}^n I_i\ib\RR^n$ is a box, we write $B_i^+$ and $B_i^-$ for
the pair of faces of $B$ whose outward normal are (respectively)
the positive and negative $i$th semi-axis.  Thus, if $I_i=[a_i,b_i]$
then $B_i^-= I_1\times \cdots \times I_{i-1}\times a_i \times I_{i+1}\times\cdots \times I_n$,
and $B_i^+$ is similar, but with $b_i$ in place of $a_i$. The center
of a box $B$, $\cen(B)$, is defined as the vector $((a_1+b_1)/2, \dd (a_n+b_n)/2)$.
For a positive real number $\lambda$, define the scaled box 
$$\lambda B \as \{ \lambda (\bfx - \cen(B) + \cen(B))| \bfx \in B\}.$$
For $\bfX \in \intbox \RR$, define the \dt{magnitude of $\bfX$}, 
$\magn(\bfX) \as \max_{x \in \bfX} \abs{x}$.

Miranda's theorem \cite{kulpa:poincare-miranda:97} gives us a sufficient condition for the existence of roots of $F$
in the interior of box $B$: 

\bproT{Simplified Miranda}{miranda}
Let  $F=(f_1\dd f_n):\RR^n\to\RR^n$ be a continuous function, and $B$ a box.
A sufficient condition that $F$ has a root in the interior of $B$ is that
	\beql{fib}
	f_i(B_i^+)>0,\qquad f_i(B_i^-)<0
	\eeql
holds for each $i=1\dd n$.
\eproT

Remark: we have stated Miranda's theorem in the simplest possible form.
For instance, our simple form could be generalized by
replacing \refeq{fib} with the following condition:
{\em $f_i$ takes a definite sign $s_i^+\in\set{-1,+1}$ on $B_i^+$,
takes a definite sign $s_i^-$ on $B_i^-$, and $s_i^+ s_i^-=-1$.}
But the simplified form implies this more general form since
we can replace the system $F=(f_1\dd f_n)$ by
	$$\wt{F}=(s_1^+ f_1\dd s_n^+ f_n),$$
since the systems $F$ and $\wt{F}$ have exactly the same set of roots.
The usual statement of Miranda's theorem is even general, where \refeq{fib} is replaced by:
{\em there exists a permutation $\pi$ of the indices $\set{1\dd n}$ with this property:
for each $i$, $f_i$ has definite signs $s_i^+$ and $s_i^-$ on $B_{\pi(i)}^+$ and $B_{\pi(i)}^-$ (respectively),
where $s_i^+ s_i^-=-1$.}
We shall see that there is no need to find such a permutation,
if we transform $F$ appropriately.
Moore and Kioustelidis \cite{moore-kioustelidis:test:80}
give the following effective form of the Miranda test:

\bproT{Effective Miranda's Test}{mk}
Let $F\as (f_1 \dd f_n):\RR^n\to\RR^n$ be a continuous function with appropriate box functions.
Write $f_{i,j} \as \partial f_i/\partial x_j$.  For any box $B$ with width $w(B)=(w_1\dd w_n)$,
if for all $i=1 \dd n$
        \beqarray
          f_i(\cen(B_i^+)) &\cdot& f_i(\cen(B_i^-))< 0, \label{eq:mk1}\\
          |f_i(\cen(B_i^+))| &>& \sum_{j=1, j\neq i}^n \magn(\intbox f_{i,j}(B_i^+)) w_j,
			\text{ and } \label{eq:mk2}\\
          |f_i(\cen(B_i^-))| &> &\sum_{j=1, j\neq i}^n \magn(\intbox f_{i,j}(B_i^-)) w_j,
			\label{eq:mk3}
        \eeqarray
then $F$ has a zero in the interior of $B$.
\eproT
\bpf
Using the mean-value interval extension of $f$, we know that
        $$f_i(B_i^+) \ib f_i(\cen(B_i^+)) + \intbox \diffop f_i(B_i^+) \cdot (B_i^+ - \cen(B_i^+));$$
note the dot-product on the RHS is the inner-product of interval vectors. But
        \begin{align*}
        \intbox \diffop f_i(B_i^+) &\cdot (B_i^+ - \cen(B_i^+)) \\
                &= \sum_{j=1}^n \intbox f_{i,j}(B_i^+) ([\ulx_j,\olx_j] - (\ulx_j+\olx_j)/2).
        \end{align*}
Since $\olx_i = \ulx_i$, the $i$th entry in the summation vanishes on the RHS and hence we obtain
\begin{align*}
        \intbox \diffop f_i(B_i^+) &\cdot (B_i^+ - \cen(B_i^+)) \\
                &= \sum_{j=1, j\neq i}^n \intbox f_{i,j}(B_i^+) ([\ulx_j,\olx_j] - (\ulx_j+\olx_j)/2)\\
                &= \sum_{j=1, j\neq i}^n \intbox f_{i,j}(B_i^+) \frac{(\olx_j-\ulx_j)}{2} [-1,1]\\
                &= \sum_{j=1, j\neq i}^n \magn(\intbox f_{i,j}(B_i^+)) \frac{(\olx_j-\ulx_j)}{2} [-1,1]\\
                &= \left(\sum_{j=1, j\neq i}^n \magn(\intbox f_{i,j}(B_i^+))\frac{(\olx_j-\ulx_j)}{2} \right) [-1,1]\\
                &= \left(\sum_{j=1, j\neq i}^n \magn(\intbox f_{i,j}(B_i^+)) (w_j/2) \right) [-1,1].
\end{align*}
Thus
        $$w( \intbox \diffop f_i(B_i^+) \cdot (B_i^+ - \cen(B_i^+)) )
                = \sum_{j=1, j\neq i}^n \magn(\intbox f_{i,j}(B_i^+))w_j.$$
Therefore, \refeq{mk2} implies that $0\nin f_i(B_i^+)$.
Similarly, \refeq{mk3} implies that $0\nin f_i(B_i^-)$.  
By \refeq{mk1}, $f_i$ takes opposite signs on the faces $B_i^+$ and $B_i^-$,
and so Miranda's theorem implies $B$ contains a root in its interior.
\epf

Miranda's test is not a ``complete'' method for detecting roots in the following sense:
there are systems $F=0$ whose roots cannot be detected by
Miranda's test, even in the general form that allows permutation $\pi$.
For instance, let $F=(f,g)$ where $f=x+y$ and $g=x-y$.
Then no rectangle $B\ib\RR^2$ containing the root $(0,0)$ will pass the generalized Miranda test.

The solution is a ``preconditioning'' trick.
Consider a transformation of $F$ to $G \as YF$, where $Y$ is a suitable non-singular matrix in the box $B$.
Note that $G$ and $F$ have the same sets of roots.  To perform the Miranda Test on a
box $B$, we choose $Y$ to be the inverse of any non-singular Jacobian $J_F(m)$ where $m\in B$.
More precisely, 
        \beql{mktest}
        \text{
	  \fbox{\begin{minipage}{2.5in}
          \dt{MK-test} for a system $F$ on a box $B$  is the effective Miranda-test applied to the 
          system $J_F(m)^{-1}F$,  where $m\as \cen(B)$, and the Jacobian is non-singular.
        \end{minipage}}}
       \eeql

This idea was first mentioned by Kioustelidis and its completeness was shown by Moore-Kioustelidis \cite{moore-kioustelidis:test:80}.
We reproduce their result, but to do that we need some notation and the Mean Value Theorem in higher
dimensions.

Given $x, y \in \RR$, the notation $x \pm y$ denotes a number of the form $x + \theta y$, 
where $\theta$ is such that $0 \le |\theta| \le 1$;
thus ``$\pm$'' hides the $\theta$ implicit in the definition. 
We further extend this notation to matrices in the following sense:
for two matrices $A, B$, the matrix $A \pm B \as [a_{ij} \pm b_{ij}]$; also, for a scalar $\lambda$,  the matrix
$A \pm \lambda \as [a_{ij} \pm \lambda]$.
We now recall the Mean Value Theorem for $F: \RR^n \mt \RR^n$: 
Given two points $\bfx, \bfy \in \RR^n$, there exists a matrix
$K$ with non-negative entries such that
        \beql{mvt}
        F(\bfx) - F(\bfy) = (J_F(\bfy) \pm K \|\bfx-\bfy\|) \cdot (\bfx -\bfy).
        \eeql
To see this claim, we apply the mean value theorem twice in each of the components of $F$ to obtain
\begin{align*}
       &f_i(\bfx) - f_i(\bfy) \\
                &\;= (f_{i,1}(\bfy) \pm K_{i,1}\|\bfx-\bfy\|, \cdots, f_{i,n}(\bfy) \pm K_{i,n} \|\bfx-\bfy\|) \cdot (\bfx-\bfy)\\
                &\;= \diffop f_i(\bfy) \cdot (\bfx-\bfy) \pm (K_{i,1} \dd K_{i,n}) \cdot (\bfx-\bfy) \|\bfx-\bfy\|  
\end{align*}
for $i=1 \dd n$.

\bleml{termination}
Let $F$ be a zero-dimensional system of polynomials. For all sufficiently small {\em open} boxes $B$
containing a single root $\alpha$ of $F$ the modified system $G \as J_F(m(X))^{-1} F$, if well defined, satisfies 
the conditions in Miranda's theorem, namely for $i=1 \dd n$, $g_i(B_i^+) \ge 0$ and $g_i(B_i^-) \le 0$.
\eleml
\bpf
Let $\bfx $ be a point on the boundary of the box $B$.
From the definition of $G$ and from the mean value theorem \refeq{mvt} we know that
\begin{align*}
  G(\bfx) &=  J_F(m)^{-1}(F(\alpha) + (J_F(m) \pm K \|\bfx-\alpha\|) \cdot (\bfx-\alpha))\\
          &= J_F(m)^{-1} (J_F(m) + K \|\bfx-\alpha\|) \cdot (\bfx-\alpha))\\
          &= (1 \pm \|J_F(m)^{-1}K\|_\infty \|\bfx-\alpha\|) \cdot (\bfx-\alpha).
  \end{align*}
The $i$th component in the vector 
        \beql{xalpha}
        (1 \pm \|J_F(m)^{-1}K\|_\infty \|\bfx-\alpha\|) \cdot (\bfx-\alpha)
        \eeql
is the polynomial $g_i(B)$, so we obtain
        \beql{gix}
        |g_i(\bfx) - (x_i - \alpha_i)| \le \|\bfx-\alpha\| \|J_F(m)^{-1}K\|_\infty \sum_{j=1}^n |x_j - \alpha_j|.
        \eeql
The term on the RHS
\begin{equation*}
        \begin{split}
          \|\bfx-\alpha\| \|J_F(m)^{-1}K\|_\infty \sum_{j=1}^n |x_j - \alpha_j| \\
                \le   \|\htw(B)\|_1^2\;\; \|J_F(m)^{-1}K\|_\infty,          
        \end{split}
\end{equation*}
because $\|\bfx-\alpha\| \le \|\htw(B)\|_2 \le \|\htw(B)\|_1$ and 
$\sum_{j=1}^n |x_j - \alpha_j| \le \|\htw(B)\|_1$.
Suppose the box $B$ is such that 
        $$2\|\htw(B)\|_1^2\;\; \|J_F(m)^{-1}K\|_\infty < \min_{i=1 \dd n} \|\alpha-B_i^\pm\|$$
then we claim that for all $i=1 \dd n$, 
$g_i(B_i^+) \ge 0$ and $g_i(B_i^-) \le 0$. This is because for all $\bfx \in B_i^+$,
$|x_i - \alpha_i|= |\olx_i - \alpha_i| = \|\alpha - B_i^+\|$, since the projection of $\alpha$ on $B_i^-$ is $(\alpha_1 \dd \alpha_{i-1}, \olx_i,\alpha_{i+1} \dd \alpha_n)$;
similar argument applies for $\bfx \in B_i^-$.
Thus the term on the RHS in \refeq{gix} is smaller than $|\olx_i - \alpha_i|/2$,
which implies that $g_i(B_i^+) \ge 0$ (we can similarly show that $g_i(B_i^-) \le 0$), and therefore
the system $G(\bfx)$ has the same sign pattern as $\bfx-\alpha$ on the boundary of the box $B$.
\epf

This ``orthogonalization'' around
the zero by the pre-conditioning step helps us avoid finding the permutation matrix in the general Miranda's
test. Note, however, that if the root is on the boundary of the box then the above proof breaks down. 

\newpage
  \section{Proofs and Details }

\textbf{Proof of \refThm{relorder}:} We will need the following lemma for the proof.

\bleml{fgorder}
If a box $B$ satisfies $\MK(B)$ and an $f$-vertex and a $g$-vertex share an edge $e$ of
$B$ then we can determine the relative order of the normalized curves $(S',T')$ along $e$.
\eleml
\bpf

Since the $\MK(B)$ test passed along $e$, we know that there are real numbers $a, b$ such
that either $a\cdot f(e) > b\cdot g(e)$ or $a\cdot f(e) < b\cdot g(e)$. 
To see this, recall that $\MK(B)$ test replaces the system $F=(f,g)^T$ by
the system $\wh{F}= J \cdot F$, where $J$ is the inverse of the Jacobian of $F$
evaluated at $\cen(B)$, and performs the Miranda test, \refPro{mk}, for $\wh{F}$.   
If $J=\mmat{a&-b\\c&d}$ and $\wh{F}=(\wh{f},\wh{g})^T$
then $\wh{f}=a\cdot f-b\cdot g$. The Miranda test
on $\wh{F}$ asserts that there is an edge $e$ for which either $\wh{f}(e)>0$  or $\wh{f}(e)<0$.  
The first inequality is equivalent to $a\cdot f(e)>b\cdot g(e)$, and
the second inequality is equivalent to $a\cdot f(e)<b\cdot g(e)$.
In the rest of the proof we assume that $a\cdot f(e)>b\cdot g(e)$; 
the analysis in the other case is same.

Neither $a$ nor $b$ can vanish, since that would imply
that either $f$ or $g$ has a constant sign on $e$, which is a contradiction as
both $f$ and $g$ have a vertex on $e$. Let $e(t)$ be a parametrization of $e$
with endpoints $e(0)$ and $e(1)$. Let $T_f\ib (0,1)$ be such that $f(e(t))=0 $ for all $t \in T_f$, and
let $t_f$ be the smallest element in $T_f$;
similarly define $T_g$ and $t_g$. Since both $f$ and $g$ change sign across $e$, we know that the cardinality
of $T_f$ and $T_g$ is odd. Any normalization $(S',T')$ of $(S,T)$ relative to $B$ will remove all but
one element from both $T_f$ and $T_g$, while maintaining the relative order of the remaining element. 
That order is the same as the order of $t_f$ and $t_g$ along $e$. Thus we want
to determine whether $t_f<t_g$ or $t_g<t_f$.
Suppose $ab>0$.   Then $f(e)> c\cdot g(e)$ for some $c>0$.  
There are two cases to consider:
\begin{mini_itemize}
\item $f(e(0))>0$:  then $f(e(t_g))> cg(e(t_g))=0$, which implies that $f$ is positive
  at $e([0,t_g])$ and so $t_f > t_g$;
\item $f(e(0))<0$: this similarly implies $t_f< t_g$.
\end{mini_itemize}
If $ab < 0$ then $g(e) > c\cdot f(e)$, for some $c>0$, and the claim follows from similar arguments. 
\epf

\myPara{Group I Patterns.}
Notice that using the sign of $f,g$ at the
corners of $B$, we can never detect these patterns.
For instance, for \refFig{patterns}(Ia), we will not detect the presence of the curve $S'$
because $f$ has the same sign on every corner of the box.
So we first show that they cannot arise.

\bleml{fgf}
Suppose box $B$ satisfies $\MK(B)$.  Then the patterns in Group I of \refFig{patterns} cannot occur.
\eleml
\bpf
Let $e$ be an edge of $B$ and
suppose $S'\cup T'$ intersect $e$ in three consecutive points $e(t_1), e(t_2), e(t_3)$ ($t_1<t_2<t_3)$
where $e(t)$ is a parametrization of $e$.  
The ``pattern'' of these intersections is the triple $(p_1,p_2, p_3)$ where $p_i\in \set{f,g}$.
For instance, if $e$ is the top edge of the box in \refFig{patterns}(Ia), then
the pattern is either $(f,g,f)$ or $(g,f,g)$. Our claim is equivalent to showing that
the intersection pattern of any three consecutive intersections of $S'\cup T'$ on
any edge $e$ of $B$ cannot be $(f,g,f)$ or $(g,f,g)$.

From \refLem{fgorder} we know that $f(e) > c \cdot g(e)$, for some $c \in \RR_{\neq 0}$; 
let us assume $c> 0$. 
Consider the $(f,g,f)$ pattern (the other pattern is similar).
Consider the sign of $g$ at the point $e(t_1-\vareps)$ and $e(t_3+\vareps)$ for sufficiently small $\vareps>0$.
Then $g$ must have different signs at these points --- this is because as we move from $e(t_1-\vareps)$
to $e(t_3+\vareps)$, the function $g$ changes sign exactly once, at $e(t_2)$.
Likewise, we see that $f$ must have the same sign at $e(t_1-\vareps)$ and $e(t_3-\vareps)$,
because as we move from $e(t_1-\vareps)$
to $e(t_3+\vareps)$, the function $f$ changes sign exactly twice, at $e(t_1)$ and $e(t_3)$.
Thus $f(e(t_1-\vareps))>g(e(t_1-\vareps))$
iff $f(e(t_1-\vareps))<g(e(t_1-\vareps))$.  This is a contradiction.
\epf

\myPara{Group II Patterns.}
Suppose $f, g$ have sign agreement on $B$.  
We can determine from these signs the two edges that contains $f$- and $g$-vertices.
Suppose $e$ is such an edge.   So there is an $f$-vertex
and a $g$-vertex on $e$, and from \refLem{fgorder} we know their relative ordering.

\myPara{Group III Patterns.}
Let us say that $f,g$ have \dt{sign agreement} on $B$ if there
is a sign $s\in\set{+1,-1}$ such that $\sign(f(c)g(c))=s$ for each corner $c$ of $B$.
Observe that Group II patterns arise precisely because $f,g$ have sign agreement;
likewise Group III patterns arise precisely because $f,g$ do not have sign agreement.
We claim that the patterns in Group III can be determined by signs
of $f$ and $g$ at the corners of $B$.
First of all, by evaluating the signs of $f$ and $g$ on the corners of $B$,
we can determine whether or not $f,g$ have sign agreement of $B$.
If not then we can determine whether the pattern is (IIIa), (IIIb) or (IIIc).
If (IIIa), the pattern is completely determined.  If (IIIb), there is an edge $e$
containing both an $f$- and a $g$-vertex, and we need to know their relative order on $e$.
This is determined by the positions of the other $f$-vertex and other $g$-vertex:
this is because the order of the four $f$- and $g$-vertices on the boundary of $B$
must be alternating: $f,g,f,g$.   A similar remark applies in case (IIIc).

To summarize the proof of \refThm{relorder}: \refLem{fgf} implies that Group I patterns cannot occur; for 
Group II patterns we can determine the relative order from \refLem{fgorder} and for
Group III patterns the ordering is immediate. 

\ignore{
\bleml{sc}
If a box $B$ satisfies $C^f_1$ predicate, and $f$ changes sign on two edges of $B$ then
$f$ cannot change sign on both the remaining two edges.
\eleml
\bpf
Let $e$ be one of the remaining edges, and suppose $f$ has a sign change on $e$.
Since $f$ has a branch corresponding to the two vertices, it follows that $f$ is not
monotone in both $x$ and $y$ direction. But this is a contradiction since $B$ satisfies
$C_1^f$ predicate.
\epf
}

\myPara{The RefineRoot Procedure:}
\progb{
{RefineRoot($B$)}\\
\Comment{Assume that $\JC(6B)$ holds.}\\
\Comment{Thus no neighbor of $2B$ can be an MK-box.}\\
 Input: an aligned box $B$ with $2B$ as the root box.\\
 Output: an aligned box $B^* \ibp 2B$ with $2B^*$ as the root box.\\
\textit{ Algorithm} $\triangleright$\\
 Remove $B$ from $Q_\MK$.\\
 Subdivide the neighbors of $B$ until the size of the \\ 
 neighborhood of $B$ is $w(B)/2$.\\
 Add the children of the neighbors to the appropriate \\
queues  $Q_0$, $Q_f$, $Q_g$, $Q_{fg}$.\\
 Initialize $Q_\tmp$ with the neighbors of $B$ and its children.\\
 While $Q_\tmp$ is non-empty do\\
\> $B_\tmp \ass Q_\tmp.pop()$.\\
\> If $\MK(2B_\tmp)$ holds then\\
\>\>  Empty $Q_\tmp$ into $Q_{fg}$. \\
 \>\> Return $B_\tmp$ and add it to $Q_\MK$ .\\
\> Else  Subdivide $B_\tmp$ and add its children to \\
\> $Q_0$, $Q_f$, $Q_g$ and $Q_\tmp$ if they satisfy\\
\> respective predicates..
}

Correctness: The subdivision of $B$ and its neighborhood of size $w(B)/2$ covers $2B$, the root box corresponding
to $B$. Let $B'$ be any of these 16 boxes. Since $\JC(6B)$ holds, if $\MK(2B')$ holds for a box $B'$ then
the root in $2B'$ is exactly the  root in $2B$. 

We now give the details of various stages mentioned in \refSec{algo}.

\myPara{Details of Stage III:}
	\progb{
	While $Q_{\JC}$ is non-empty\\
	\>	$B\ass Q_{\JC}.pop()$.\\
	\>	$Q_\tmp\ass \set{B}$\\
	\>	While $Q_\tmp$ is non-empty do\\
        \>\>                $B_\tmp \ass Q_\tmp.pop()$.\\
	\>\>		If $\MK(2B_\tmp)$ holds.\\
	\>\>\>		  Push $B_\tmp$ into $Q_\MK$.  Empty $Q_\tmp$ into $Q_{fg}$.\\
	\>\>	Else  subdivide $B_\tmp$ and distribute the children \\
        \>\>            into $Q_0, Q_f, Q_g, Q_\tmp$ \\
        \>\>            (after testing for the corresponding predicates).\\
        For each box $B \in Q_\MK$ do\\
        \> If there is another box $B' \in Q_\MK$ such that \\
	\> $2B \cap 2B' \neq \es$ then remove $B'$ from $Q_\MK$. 
	}
        Note that we only search for a root in $fg$-candidate boxes. This is justified by \refLem{termination}
        and the observation that eventually the root will be contained in the interior of the doubling
        of an $fg$-candidate box.
	At the end, $Q_\JC$ is empty and $Q_\MK$ contains a set of root boxes. Moreover,
        the last loop ensures no two boxes $B, B' \in Q_\MK$ correspond to the same root, i.e., $2B \cap 2B' =\es$.
        The boxes in $Q_{fg}$  do not contain any root. 

\myPara{Details of Stage V:} For each box $B$ in $Q_\Root$ do the following steps.
        \progb{
          Initialize $Q_\tmp$ with all the neighbors of $B$ in $\TT$.\\
          While $Q_\tmp$ is non-empty do\\
          \> $B_\tmp \ass Q_\tmp.pop()$.\\
          \> If $B_\tmp \ibp 8B$ then turn it OFF and\\
          \> add its neighbors to $Q_\tmp$.\\
          \> If the interior of $B_\tmp$ intersects the interior of $8B$ then\\
	  \> subdivide it and add its children to $Q_\tmp$.\\

          \Comment{NOTE: Whenever we subdivide a box $B_\tmp$}\\
	  \Comment{we remove it from one of the queues $Q_f$, $Q_g$, or $Q_{fg}$}\\
	  \Comment{and add its children to the appropriate queue.}
          }
          Since $8B$ is half-aligned, there is a refinement of $\TT$ such that every box
          in this refinement is either contained in $8B$ or does not intersect its interior. Thus the 
          procedure described above will terminate. Let $\TT'$ be the refinement of $\TT$ with blacked-out
          regions corresponding to extended root boxes. 

\myPara{Details of Stage VI:}
        \progb{
          For each $B \in Q_\Root$ do\\
          \> Let $m$ be the largest depth amongst all the \\
	  \> neighbors $B_\tmp$ of $8B$ in $\TT'$.\\
          \> Let $\ell$ be the depth of $B$ in the subdivision tree $\TT$.\\
	  \> \Comment{Thus $w(B_\tmp) = w(B)2^{\ell-m}$}\\
          \> If $m > \ell$ then $k \ass m$; else $k \ass \ell + 1$.\\
          \> Add a {\em conceptual leaf box} to $\TT'$ that represents $8B$.\\
	  \> Set the depth of this box to $k+1$ and  initialize\\
          \> its 8 pointers to the 8 neighbors of $8B$ in $\TT'$.\\
          Let $\TT''$ be the resulting subdivision tree.\\
          Let $Q$ be the priority queue of all the leaves in $\TT''$;\\
	  the deeper the level the higher the priority.\\
          Initiate the standard balancing procedure \\
	  on $Q$ with one difference: whenever we pop a conceptual\\
	  box $8B$  we check the depth of its neighbors and if\\ 
          necessary reset the depth of $8B$ to one more than\\
          the depth of its deepest neighbor.\\
          \Comment{NOTE: Whenever we subdivide a box $B_\tmp$ we}\\
	  \Comment{ remove it from one of the queues $Q_f$, $Q_g$, or $Q_{fg}$}\\
	  \Comment{and add its children to the appropriate queue.}
        }
        We claim that at the end of this procedure the tree $\TT'$ is balanced, and all the neighbors
        of extended root boxes $8B_i$ in $\TT'$ are of the same size, namely $w(B_i)/2^k$, for some $k \ge 1$. 
        The balancing of $B_0 \setminus \cup_i (8B_i)$ follows from the proof of correctness for 
        standard balancing procedure. The conformity follows because 
        a conceptual box is always deeper in $\TT''$ than its neighbors, so it will never be subdivided, and
        its neighbors will always be twice its size.
        The modification to the standard balancing is required, because a smallest neighbor $B_\tmp$ of $8B$
        in $\TT'$ could have been subdivided by a box that is adjacent to $B_\tmp$ along the edge that
        is not abutting $8B$ or any of the neighbors of $8B$. However, this can only happen once
        because of the balancing property, \refLem{balance}.

\myPara{Details of Stage IX:}
       \progb{
          Initialize $Q_\tmp$ with all the root boxes.\\          
          \Comment{$Q_\tmp$ will contain unambiguous boxes.}\\
          For each box $B \in Q_{fg}$ do\\
           If there is pair of $f$-vertex and $g$-vertex that do not share\\ a segment of $B$ then \\
          \> Connect the two $f$-vertices with an edge;\\
          \> connect the two $g$-vertices with an edge; \\
          \> ensure that the two edges do not intersect. \\
         \> Add $B$ to $Q_\tmp$ and remove it from $Q_{fg}$.\\
           \Comment{In the remaining boxes, the two pairs of }\\
	   \Comment{$(f,g)$-vertices share the same segments.}\\
           If the two pairs of $(f,g)$-vertices are on edges $e$, $e'$ \\
           that share a vertex then \\
          \>  \Comment{Call such a box a \dt{Transition Box}}\\
          \> \Comment{These boxes definitely appear in a covering}\\
	  \> \Comment{of nested $fg$-loops; they can appear otherwise.}\\
          \> Subdivide both $e$ and $e'$ until we reach\\ 
	  \> a subset $e''$ in one of the edges such that\\
          \> only one of the curves $f$ or $g$ changes sign on $e''$;\\
	  \> say $e'' \ibp e$ and $f$ changes sign on it.\\
          \> Check which side of $e\setminus e''$ does $g$ change sign; \\
	  \> order the $f$-vertex and $g$-vertex\\
          \> along $e$ accordingly; connect the $f$-vertices \\
	  \> and $g$-vertices respecting this order;\\
          \> add $B$ to $Q_\tmp$ and remove it from $Q_{fg}$.\\
          If $B$ shares an edge $e$ with $B_0$ then\\
         \> Subdivide $e$ until we reach a subset $e'' \ibp e$ \\
	 \> such that only one of the curves $f$, $g$ changes\\
         \> sign on $e''$. Check which side of $e\setminus e''$ contains\\
	 \> the other curve. Order the vertices accordingly\\
         \> and connect the $f$-vertices and $g$-vertices. \\
	 \> Add $B$ to $Q_\tmp$ and remove it from $Q_{fg}$.\\
         \Comment{The boxes in $Q_\tmp$ are all unambiguous boxes.}\\
          While $Q_\tmp$ is non-empty do \\
           $B \ass Q_\tmp.pop()$\\
           For each ambiguous $fg$-candidate $B_\tmp$ of $B$ do\\
          \> Order the $f$-vertices and $g$-vertices on the \\
	  \> shared segment between $B_\tmp$ and $B$ according\\
          \> to their ordering in $B$; connect the pair of $f$-vertices\\
	  \> and $g$-vertices in $B_\tmp$ respecting this ordering.\\
          \> Add $B_\tmp$ to $Q_\tmp$ and remove it from $Q_{fg}$.\\
	  \> \Comment{Thus all the $fg$-neighbors are unambiguous.}
         }
In practice, we should first resolve boxes that can be traced to root boxes.
Then we should resolve transition boxes
and propagate their ordering. Finally, in the remaining ambiguous boxes, we should resolve the
boundary boxes and propagate their ordering. At the end of this stage $Q_{fg}$ will be empty,
since any ambiguous box can be traced to one of the four boxes: root box, transition box,
or a boundary box.


\end{appendices}


\end{document}